\documentclass{ws-procs9x6}
\usepackage{times}
\begin{document}

\title{Surprises in the RHIC Data\footnote{\uppercase{F}or additional
information see http://www.star.bnl.gov, http://www.phenix.bnl.gov,
http://www.phobos.bnl.gov, and http://www.rhic.bnl.gov/brahms. }}

\author{J.H. Thomas\footnote{\uppercase{W}ork  supported by the Office
of \uppercase{S}cience at the \uppercase{US} \uppercase{D}epartment of
\uppercase{E}nergy.}}

\address{Lawrence Berkeley National Laboratory \\
B510A, 1 Cyclotron Rd., \\
Berkeley, CA, USA \\
E-mail: jhthomas@lbl.gov }

\maketitle

\abstracts{  
The data  from RHIC have  produced many unanticipated results.  I will
describe a few  of the surprises that occur in  the soft spectra while
my colleagues at this conference  will summarize the hard spectra. One
particularly  important discovery  is that  properties of  the initial
state  have an impact  on the  final state  in relativistic  heavy ion
collisions.  Another important discovery is that the collision zone is
opaque  to the  passage  of  hadrons and  perhaps  even partons.   And
finally, the data  tell us very precisely where  the colliding systems
hadronize on the phase diagram for nuclear matter. }

\section{Introduction }

The Relativistic  Heavy Ion Collider  (RHIC) is located  at Brookhaven
National Laboratory on Long Island,  New York. The collider is 3.83 km
in circumference and it accelerates a variety of heavy ion beams; from
Au to protons. The top energy is  100 GeV/amu per beam for Au ions and
250 GeV  per beam for protons.  The top collision  energies are $\surd
$s$_{NN}$ = 200 GeV and $\surd $s$_{pp}$ = 500 GeV.

In  this paper,  I will  summarize the  results recorded  in  the soft
spectra (p$_{T}$  $\leq$ 1 GeV)  that were observed during  the $\surd
$s$_{NN}$  =  130  and  $\surd  $s$_{NN}$  = 200  GeV  Au-Au  runs  at
RHIC. There  are many unanticipated results  in these data  and I will
focus on those  things that I have found to  be surprising relative to
our expectations when we started the construction of the accelerator.

\begin{figure}[htbp]
\centerline{\epsfysize=3in\epsfbox{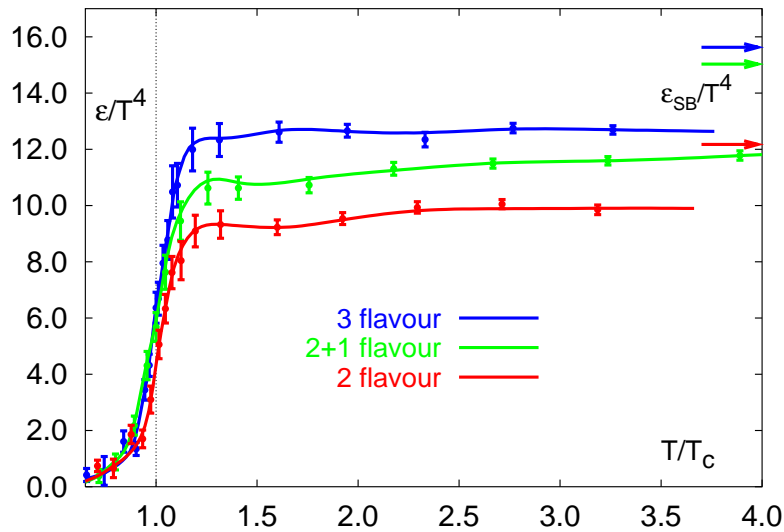}}
\caption{Lattice gauge calculation by  F. Karsch of the energy density
in a  system of quarks with  2 or 3  flavors. The arrows on  the right
hand side of the figure indicate the Stephan Boltzman limit for a free
Quark Gluon gas.}
\label{fig1}
\end{figure}

The motivation  for building  RHIC was to  study nuclear  matter under
extreme  conditions; at high  temperature and  at high  density. Under
these  conditions, we  expect quark  and gluon  degrees of  freedom to
become  important and  the underlying  dynamics should  change  as the
nuclear system makes the transition  from cold matter to extremely hot
and dense matter.  In fact, it was predicted  that nuclear matter will
undergo  a phase  transition  into a  Quark  Gluon Plasma  (QGP) at  a
critical temperature  near the rest mass  of the pion and  at about 10
times the  density of  normal nuclear matter.  This scenario  has been
explored with lattice gauge calculations\cite{Karsch:1} and the
calculations predict  that there  will be a  large jump in  the energy
density for  two and three  flavor systems at a  critical temperature,
T$_{c}$, of about 160 MeV, see Figure \ref{fig1}. This prediction is
remarkably stable  with respect to  changes in the  underlying lattice
technology  and  over   time.  The  state  of  the   art  for  lattice
calculations is  such that we do  not know if the  phase transition is
first order, second  order, or whether or not  there is a tri-critical
point on the phase diagram.

\section{Surprises in the Soft Spectra at RHIC}

The  first surprise  to be  seen in  the RHIC  data is  that  the mean
multiplicity  of particles per  event is  large but  not exceptionally
large.     The     PHOBOS     collaboration     made     the     first
determination\cite{Back:2000}$^{,}$\cite{Back:2002}   of  the  maximum
multiplicity in  central collisions of  Au ions at $\surd  $s$_{NN}$ =
130 GeV and they  found a mean multiplicity of 4200 $\pm  $ 470 in the
top 3{\%}  most central collisions.  They have also measured  the mean
multiplicity at  $\surd $s$_{NN}$ = 56  GeV and at  $\surd $s$_{NN}$ =
200 GeV and, in general, the values do not suggest a large jump in the
multiplicity   of    particles   relative   to    the   other   energy
points. Instead,  the multiplicities change smoothly as  a function of
$\surd $s and  they are in fairly good  agreement with the predictions
of  the HIJING  model\cite{Wang:2001}. This  is  a bit  of a  surprise
because HIJING  was created to  represent the jets and  mini-jets that
are  formed in  heavy ion  collisions due  to the  interaction  of the
partons in the system. HIJING was  not designed to be a complete model
of  relativistic  heavy ion  collisions.  Models  with more  ambitious
designs and which  include a detailed description of  the final state,
such  as RQMD and  UrQMD\cite{Bass:1998}, are  not very  successful at
describing   the  multiplicity   of  particles   and   their  rapidity
distributions at RHIC.

\begin{figure}[htbp]
\centerline{\includegraphics[height=3in]{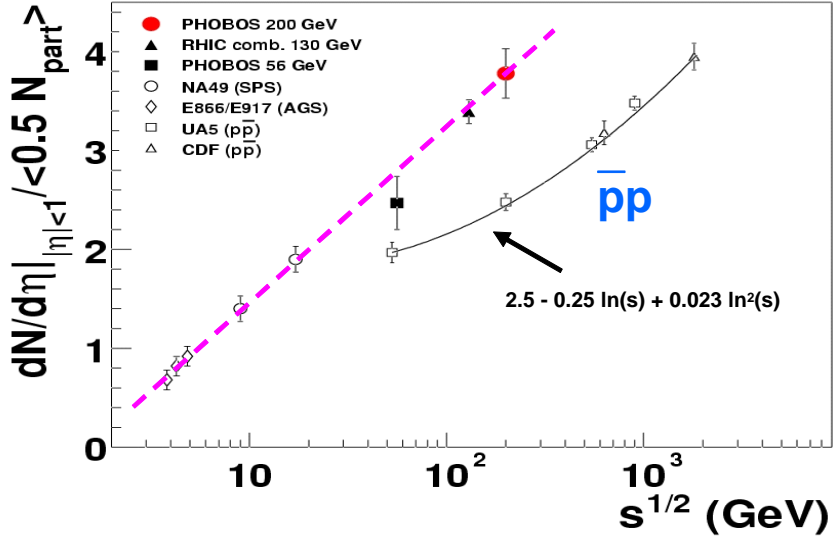}}
\caption{Particle yields  per participating  nucleon  increase with
beam energy as shown in this figure from the PHOBOS collaboration. The
rate of increase is predicted by  the model of Kharzeev at al. however
the lines on the figure are merely to guide the eye.}
\label{fig2}
\end{figure}

A  better description of  particle multiplicities  was first  given by
Kharzeev   and   Nardi\cite{Kharzeev:2001}   and   by   Kharzeev   and
Levin\cite{Kharzeev:2002}. Their theme,  however, is that the particle
multiplicities are determined by  the properties of the incoming state
and  not  by  the  dynamics   of  the  final  state.  They  and  their
collaborators have  proposed that the  gluon spectrum in  the incoming
state is  modified by the Lorentz  contraction of the  nucleus and the
running  of the  coupling  constant so  that  the interaction  becomes
coherent  at RHIC  energies  and the  gluon interaction  cross-section
saturates (i.e.  $\rho \cdot \sigma $  = 1.0). This gives  rise to a
$\surd $s  dependent gluon  spectrum that evolves  slowly and  it gets
harder  as $\surd  $s increases.  The  increase can  be predicted  and
translated  into  particle  yield  as  a function  of  the  number  of
participating  nucleons\cite{Kharzeev:2001}  or  the  center  of  mass
energy of the collisions\cite{Kharzeev:2002}. See Figure \ref{fig2}.

Another  observation due  to  the initial  state  saturation model  of
Kharzeev and Nardi is that  the production of particles increases more
rapidly than  participant scaling.  They claim that  RHIC multiplicity
data suggest an  admixture of soft and hard  collisions and that about
15{\%} of the collisions are  hard binary collisions. The trend in the
data  is consistent  with  their initial  state  saturation model  and
inconsistent with  most final state saturation models.  The same trend
can  be seen in  the measurements  of the  total transverse  energy in
Au-Au collisions;  E$_{T}$ increases more  rapidly than the  number of
participating nucleons and requires  a 20{\%} admixture of hard binary
collisions  to explain  the  data. This  ratio  is consistent,  within
errors, with the ratio derived from the multiplicity data.

The  total transverse  energy per  rapidity  interval can  be used  to
estimate  the thermalized  energy density  in the  collision  zone via
Bjorken's formula:

\noindent
\begin{equation}
\varepsilon \; = \;\frac{1}{\pi \,R^2}\,\frac{1}{\tau _0 }\,\frac{dE_T 
}{dy}
\label{eq:one}
\end{equation}

$\tau _{0}$ is the time required  to thermalize the system and we take
it to be 1  fm/c, although it is probably smaller. R  is the radius of
the Au nucleus, and dE$_{T}$/dy  is taken from measurements. Using the
measurements by  the PHENIX collaboration\cite{Adcox:2001}  it is easy
to show that energy density is at least 4.6 GeV/fm$^{3}$ at RHIC which
is 30 times  higher than normal nuclear matter densities  and 1.5 to 2
times higher than achieved at any other accelerator.

\begin{figure}[htbp]
\centerline{\includegraphics[height=2.75in]{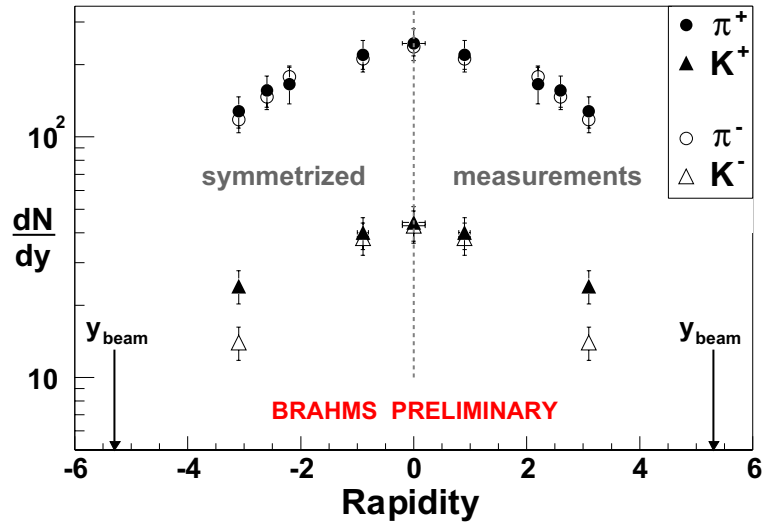}}
\caption{Pion and  Kaon spectra as a function  of rapidity at $\surd
$s$_{NN}$  =  200   GeV.  The  data  were  collected   by  the  Brahms
collaboration. The curves would be flat  out to 6 units of rapidity if
boost invariance was strictly true.}
\label{fig3}
\end{figure}

Bjorken  hypothesized that the  collision zone  is boost  invariant in
order  to  derive equation  1.   However, this  turns  out  not to  be
true. Boost invariance is approximately  valid to within $\pm $2 units
of  rapidity,  as can  be  seen in  Fig.  \ref{fig3},  but then  boost
invariance   is   incomplete  at   higher   rapidities.   The   Brahms
collaboration\cite{Ouerdane:2003}   has  measured   many   species  of
identified particles  over a wide  range of rapidities and  it appears
that boost  invariance holding  out to 2  units is a  fairly universal
feature  and it  is independent  of particle  ID.  The  observation of
incomplete  boost  invariance is  a  surprise  relative  to our  early
hypotheses  but  in  retrospect  it  was not  unexpected.   The  early
hypotheses were  deliberately simplistic  and heavy ion  reactions are
rich and complex with a large diversity of features.

\begin{figure}[htbp]
\centerline{\includegraphics[height=3.5in]{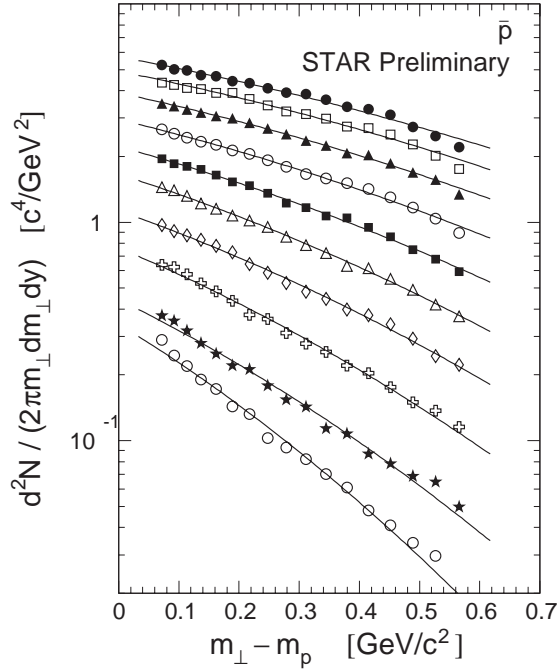}}
\caption{Anti-proton  spectra measured  by the  STAR  collaboration at
$\surd $s = 200 GeV. The lines to guide the eye are Gaussian curves of
the  form A$  \cdot $exp(-p$_{T}^{2}$/2$\sigma  ^{2})$.  The different
data   sets   represent    different   impact   parameters   for   the
collisions. The most central collisions have the largest yield.}
\label{fig4}
\end{figure}

The complexity of heavy ion reactions is shown clearly by the spectrum
of particles  that are  observed at RHIC.   Fig.  \ref{fig4}  shows an
anti-proton    spectrum    which   was    observed    by   the    STAR
collaboration\cite{Baranakova:2003}. The  spectrum is not  the Maxwell
Boltzman distribution you would  expect for massless particles because
the mass of the particles alter the kinematics of the radial expansion
of the fireball that is created  in a collision.  In the limited range
of m$_T$-m$_0$ shown in the figure,  the best fit to the spectrum is a
Gaussian.  Knowing  the effective shape  of the spectrum  is important
because we can't measure the yield of particles everywhere and we need
to  extrapolate the yields  into the  unmeasured corners  of parameter
space in order to estimate the total cross-section.

A huge  number or spectra have been  recorded at RHIC. What  can we do
with  them?  One  interesting  exercise  is to  compare  the ratio  of
particles   to    anti-particles.    The   STAR    collaboration   has
measured\cite{Cebra:2002} the  \={p}/p ratio at $\surd  $s$_{NN}$ = 20
GeV,  130 GeV,  and  200 GeV.  The  ratios are  0.11,  0.71, and  0.80
respectively or, in other words,  the ratio approaches unity as $\surd
$s increases.  And since  the anti-particle to  particle ratio  in the
early universe was 1.0, RHIC collisions are in some way similar to and
approaching the conditions in the early universe.

One  explanation  for  the  high  yield  of  anti-particles  in  Au-Au
collisions is that  they were produced by pair  production. It is easy
to show that  the anti-particle to particle ratio  of 0.8 quoted above
suggests that 80{\%}  of the protons were produced  by pair production
and  20{\%} were  carried  in by  the  beam. It  also  means that  the
mid-rapidity  region  is  not  baryon  free.   This  is  an  important
observation because many  of the early models of  heavy ion collisions
disagreed  on   the  net-baryon  number  at   mid-rapidity  and  these
observations helped weed out the unsatisfactory models.

\begin{figure}[htbp]
\centerline{\epsfxsize=4.4in\epsfysize=2in\epsfbox{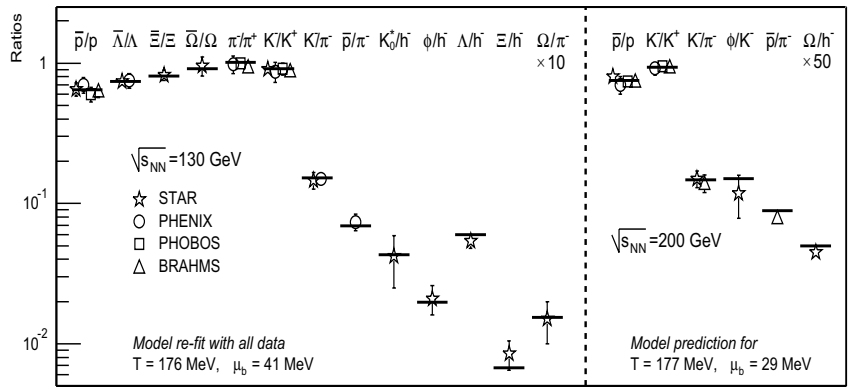}}
\caption{Particle  ratios measured  by the four  RHIC collaborations
are  compared to  the thermal  fire-ball model  of  Braun-Munzinger et
al. The agreement  between the data and the model is  very good at 130
GeV and 200 GeV.}
\label{fig5}
\end{figure}

Another  interesting exercise  is  to compare  the  ratio of  produced
particles   to  the   predictions  of   a  thermally   and  chemically
equilibrated   fireball   model.   This   has   been   done  by   many
authors\cite{Braun:2001}, but  one previously published  piece of work
was recently updated by  D.  Magestro\cite{Magestro:2002} at QM2002 to
include the most  recent 200 GeV data from  RHIC. See Fig. \ref{fig5}.
He  showed  that  the  data  are consistent  with  a  baryon  chemical
potential of 29 $\pm $ 6  MeV and a temperature for chemical freezeout
of 177  $\pm $ 7 MeV.  Chemical freeze-out marks the  end of inelastic
collisions in  a fireball. These  numbers are modestly  different than
the  values derived  from the  130 GeV  data at  RHIC where  $\mu _{B}
\approx  $  40  MeV  and  T$_{ch}  \approx $  175  MeV  and  they  are
substantially different  than the  values at the  SPS where  $\mu _{B}
\approx $ 270 MeV and T$_{ch } \approx $ 165 MeV. The trend is for the
chemical  potential to  decrease  as  a function  of  $\surd $s  while
T$_{ch}$ increases to an asymptotic value of about 175 MeV. This trend
was   recognized   in   the   low   energy  data   by   Cleymans   and
Redlich\cite{Cleymans:1998}  and  it  continues  to be  true  at  RHIC
energies.  And for whatever  it might mean, the asymptotic temperature
for  chemical freezeout is  remarkably close  to the  phase transition
energy predicted\cite{Karsch:1}$^{,}$\cite{Karsch:2002} by lattice QCD
with 2 flavors.

These results tell us very precisely where we are on the phase diagram
for nuclear matter at the  time of hadronization.  Since we know where
we are, the  challenge to the theorists is to  predict what else might
be on the  phase diagram such as the location  of a tri-critical point
or another feature that is not directly accessible by experiment.

There is  another important temperature parameter that  we can measure
and it  is the temperature that  marks the end  of elastic collisions,
T$_{kinetic}$.   It  is  lower   than  the  temperature  for  chemical
freeze-out, and below this temperature the particle momenta are frozen
and thereafter  the composition  and the kinetic  energy of  the final
state is well defined. T$_{kinetic}$ can be estimated from the inverse
slopes  of the transverse  momentum spectra  shown in  Fig. \ref{fig4}
because, in  general, all  particles are undergoing  transverse radial
expansion with  the same  expansion velocity distribution  and because
the mass of  the particles affect the shape of the  spectra due to the
different kinetic  energies involved  in their propagation.   The more
massive particles have  a larger inverse slope which  is equivalent to
saying that they  have a higher effective temperature.   The data have
been analyzed to show that there is a universal freeze-out temperature
for   all   particles   at   RHIC   and  it   is   approximately   100
MeV\cite{Kaneta:2002}. The radially expanding shock wave travels at an
average velocity  of 55{\%} to  60{\%} of the  speed of light  and the
leading  edge   travels  even  faster  (assuming   a  linear  velocity
profile). This  suggests that there is  explosive transverse expansion
of hadronic  matter after  a RHIC collision  and this  rapid expansion
generates very high pressure gradients inside the collision zone.

\begin{figure}[htbp]
\centerline{\includegraphics[height=2.2in]{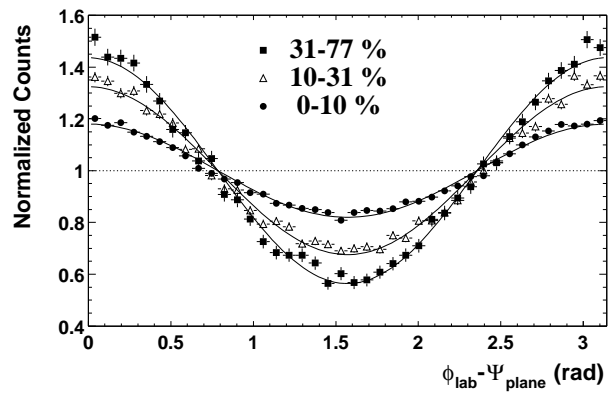}}
\caption{ Azimuthal  distributions with respect to  the reaction plane
of charged particles within 2 $\leq$ p$_{T}$ $\leq$ 6 GeV/c, for three
collision centralities.  The percentages are given with respect to the
geometrical cross  section $\sigma_{geo}$.   Solid lines show  fits to
the equation 1+2v$_{2}$cos2($\phi_{lab}  - \Psi_{plane}$).  The figure
is from Adler \textit{et al}. 2003.
 }
\label{fig6}
\end{figure}

One consequence of  the large pressure gradients is  that the emission
pattern  of final  state  particles  in the  transverse  plane is  not
isotropic. It comes about because the initial state has a well defined
anisotropy in coordinate space due  to the almond shape of the overlap
zone  when  two spherical  heavy  ions  collide  with non-zero  impact
parameter. The  anistropy in  coordinate space can  carry over  to the
momentum distribution of the final state particles if the constituents
interact early in the  collision history. These interactions build the
pressure  gradients that  drive the  flow  of particles  in the  final
state.  (Or perhaps  the converse  is more  obvious: if  there  are no
interactions amongst the initial  state constituents then the emerging
pattern of  final state particles  will be azimuthially  isotropic. So
interactions early in the collision  history are required if the final
state  particle distributions  are  observed to  be anisotropic.)  See
Fig. \ref{fig6}.

\begin{figure}[htbp]
\centerline{\includegraphics[height=2.35in]{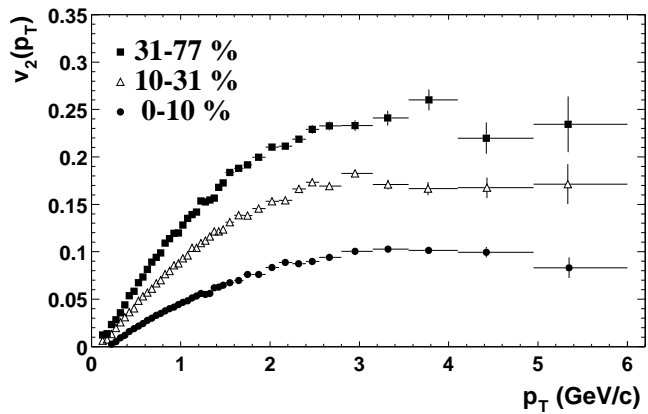}}
\caption{v$_{2}$(p$_{T}$) for different collisions centralities. The
figure is from Adler \textit{etal}. 2003.
}
\label{fig7}
\end{figure}

Figures  \ref{fig6}  and   \ref{fig7}  illustrate  the  anisotropy  of
particles in the transverse plane. This is sometimes called ``elliptic
flow''. The magnitude  of the anisotropy, as measured  by the 2$^{nd}$
Fourier coefficient  v$_{2}$, is large.   It is biggest  in peripheral
collisions\cite{Ackerman:2001}  (i.e. large  impact parameter)  and it
decreases as the impact parameter decreases. The data are in very good
agreement   with    the   predictions   of    several   hydrodynamical
models\cite{Teaney:2002}$^{,}$\cite{Huovinen:2002}  and this  is worth
noting because  the models  assume thermodynamic equilibrium  at early
times followed by hydrodynamic expansion; so thermodynamic equilibrium
is not inconsistent with the data we see at RHIC.

A surprising  feature of the  RHIC data is  that the magnitude  of the
flow signal does not decrease at high p$_{T}$ although we expect it to
decrease as more and more energetic particles pop out of the collision
zone. We observe  that the magnitude of the  elliptic flow is constant
to    the    highest    p$_{T}$    which   we    can    measure    (12
GeV)\cite{Filimonov:2003}  and this  suggests that  there  are unusual
energy loss mechanisms that cause even the most energetic particles to
interact at early times in the collision history.

The large amount of elliptic  flow suggests that the collision zone is
not  transparent to  the  passage  of hadrons  and  partons. There  is
additional evidence for this lack  of transparency, or opacity, and it
comes from the analysis of Hanbury-Brown-Twiss correlations.

\begin{figure}[htbp]
\centerline{\includegraphics[height=1.6in]{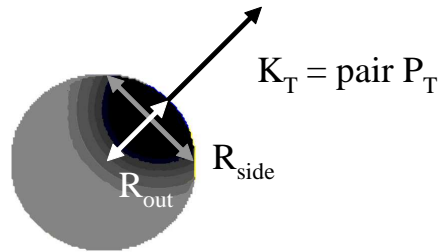}}
\caption{The  R$_{out}$,  R$_{long}$,  R$_{side}$  coordinate  system.
R$_{long}$ goes  into the page.   The sphere represents  the collision
zone at the time of hadronic freeze-out.}
\label{fig8}
\end{figure}

HBT is a method for  observing pairs of particles and the correlations
in the spectra can reveal the  size of the source that is emitting the
particles.  The quickest  explanation of the technique is  to say that
pions  undergo Bose  condensation at  the  point of  creation and  the
resulting correlations in phase space cause the pions to be correlated
in coordinate space before and  after thermal freeze-out. It will turn
out that the Fourier transform  of the momentum correlation is related
to  the radius  of the  source. The  usual coordinate  system  for HBT
analysis is  R$_{long}$, R$_{out}$, and  R$_{side}$.  It is a  pair by
pair     and    event    by     event    coordinate     system.    See
Fig.  \ref{fig8}.  R$_{long}$ is  the  radius  of  the source  in  the
direction  of  the  Z axis  (usually  chosen  to  lie along  the  beam
direction). In  our example, R$_{long}$ goes into  the page. R$_{out}$
is the radius of the source in the direction of the summed momentum of
the  pair, K$_{T}$.  R$_{side }$is  the radius  of the  source  in the
direction transverse  to both the  Z axis and K$_{T}$.   R$_{side }$is
the parameter  most easily  related to the  geometry of  the collision
zone. It is relativistically invariant and it represents the geometric
radius  of  the  system  in  the  transverse  plane  (neglecting  flow
effects).  R$_{out}$ is  more complex  and it  involves  the geometric
radius of  the system  as well  as the relative  velocity and  time of
emission of  the two particles  because the particles will  be further
separated in space at the  detector if the velocities and times aren't
identical when they are produced.

Figure 9  shows a summary of  the data for several  experiments at the
AGS, the  SPS, and  at RHIC.  In  general, R$_{long}$,  R$_{out}$, and
R$_{side}$ are very  similar at all energies however  a careful study
will reveal  that R$_{long}$  evolves slowly as  a function  of $\surd
$s$_{NN}$ and there are no unusual changes in R$_{side}$ and R$_{out}$
at RHIC  energies. These later two  radii are essentially  the same at
all energies.

\begin{figure}[htbp]
\centerline{\includegraphics[height=3.4in]{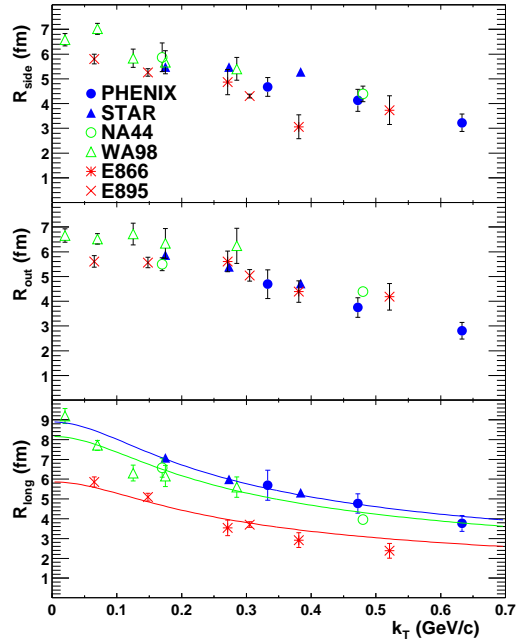}}
\caption{HBT radii for pion pairs as a function of k$_{T}$ measured at
midrapidity  for various  energies  from E895  ($\surd$s$_{NN}$ =  4.1
GeV), E866 ($\surd$s$_{NN}$ = 4.9 GeV), NA44 and WA98 ($\surd$s$_{NN}$
= 17.3 GeV), and STAR and  PHENIX at RHIC ($\surd$s$_{NN}$ = 130 GeV).
The  bottom plot  includes fits  to A/$\surd$m$_{T}$  for  each energy
region.  The figure is from Adcox \textit{et al}. 2002.}
\label{fig9}
\end{figure}

The conventional  wisdom (before RHIC)  was that the  entire collision
zone would  emit particles and R$_{side}$ would  measure the geometric
radius of the  collision zone.  In this scenario,  R$_{out}$ is always
greater  than R$_{side}$  because R$^2_{out}  \approx $  R$^2_{side} +
\beta^2 \tau^2$ where $\tau$ is the duration of particle emission.  We
expected the ratio of R$_{out}$/R$_{side}$ to be greater than one, and
perhaps  much  larger than  one,  due to  the  long  delay in  forming
particles as the system loses entropy after an energetic collision.

\begin{figure}[htbp]
\centerline{\includegraphics[height=3.5in]{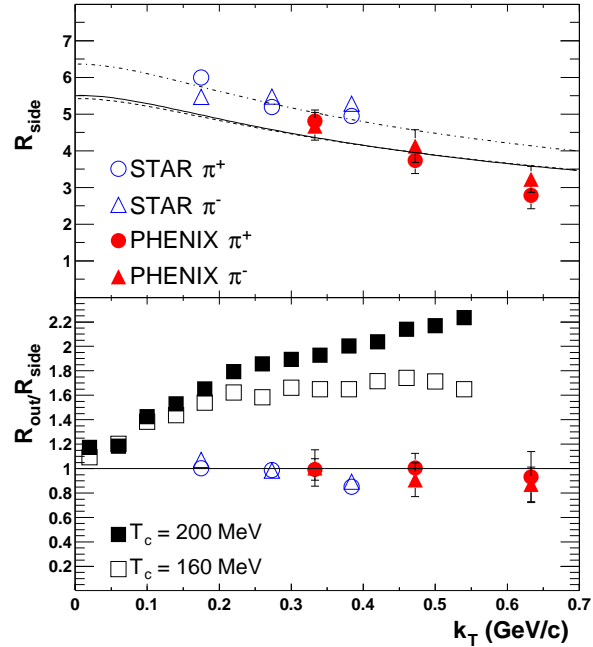}}
\caption{The top  panel shows  the measured R$_{side}$  from identical
pions at STAR  and PHENIX.  The dot and dashed  lines are explained in
Adcox  \textit{et  al}.  2002.   The  bottom  panel  shows  the  ratio
R$_{out}$/R$_{side}$   as  a  function   of  k$_{T}$   overlayed  with
theoretical  predictions  for  a  phase  transition  at  two  possible
critical temperatures.}
\label{fig10}
\end{figure}

This has not  turned out to be the  case at RHIC. R$_{out}$/R$_{side}$
is  $ \approx  $  1.0 and  the ratio  is  constant, or  falling, as  a
function of k$_{T}$\cite{Adler:2001}$^{,}$\cite{Adcox:2002}.  See Fig.
\ref{fig10}.  This was  a surprise  and it  has been  named  ``the HBT
puzzle''.   A possible explanation  for this  observation is  that the
collision  zone is opaque  and the  full depth  of the  collision zone
can't  emit  particles that  reach  the  detectors.   Instead, only  a
restricted  zone near  the  surface actually  emits  particles in  the
direction normal to  the surface and this allows  R$_{out}$ to be very
thin, indeed.  This interpretation  is suggested by the darker shading
and by the length of the arrows in Fig. \ref{fig8}.

The  diagram in  Fig.  \ref{fig8}  is  more than  a sketch.   It is  a
calculation of the where the  pions are emitted according to the Blast
Wave  Model\cite{Lisa:1}.  The Blast  Wave is  a hydro  inspired model
that attempts to  describe the particle spectra at  RHIC including the
shape  and mass  dependence of  the spectra,  it describes  radial and
elliptic flow, and  it describes HBT. It is  not a fundamental theory,
it is an effective theory, but  it was discovered that the model needs
a  parameter to  define  the size  of  the transparent  sector of  the
collision zone and this parameter must be less than the full radius of
the source in order to properly describe all of the available data.

\section{Summary}

Nuclear matter  at RHIC is very  surprising. It is hot,  its fast, its
opaque   and  yet   its  properties   still  remain   consistent  with
thermodynamic  equilibrium.  Inelastic   collisions  freeze-out  at  a
temperature  of   175  MeV.    Elastic  collisions  freeze-out   at  a
temperature  of 100  MeV. In  addition, the  radially  expanding shock
front that  is produced by heavy  ion collisions is  traveling at more
than  55{\%}  of  the speed  of  light.  There  are large  amounts  of
anisotropic transverse flow in  the collision zone. This suggests that
the  system is  undergoing  hydrodynamic expansion  due  to very  high
pressure gradients developed early  in the collision history. Finally,
the  collision zone  is not  fully transparent  and this  disrupts HBT
correlations and, as you will see in other talks in these proceedings,
the  lack of  transparency implied  by these  results extends  to high
p$_{T}$ phenomena as well.

\end{document}